\documentclass{article}
\usepackage{graphicx} 
\usepackage{amsmath, amssymb, float, enumerate, color, hyperref}
\usepackage[table,xcdraw]{xcolor}       
\usepackage{dcolumn}    
\usepackage{tabularx}
\usepackage{caption, subcaption} 
\newcolumntype{d}[1]{D{.}{\cdot}{#1} }
\title{Computational Redistricting of Iowa's Congressional Districts}
\author{Stefanie Wang and Nathaniel Merrill }
\date{July 2025}

\begin{document}

\maketitle

\abstract{This article expands on the redistricting algorithm proposed by Chen and Rodden \cite{Cutting} for states with fewer than eight congressional districts, populations highly concentrated in urban areas, or state laws that require preservation of county lines. We used the updated algorithm to redistrict Iowa's four congressional districts. Non-partisan, randomly drawn maps were used to evaluate the fairness of the enacted congressional map. Notably, the evidence suggests that the first proposed map drawn by the Iowa Legislative Bureau was rejected due to partisan bias. This article also analyzes Iowa's 2024 election results for partisan gerrymandering.


}

\section{Introduction}\label{S:intro}


In this paper, we expand on existing literature and propose methodology for evaluating whether a particular proposed legislative map is ``fair," using a case study of Congressional redistricting in the state of Iowa. 

With every decennial census, seats in the House of Representatives are reallocated from state to state based on updated population counts. Each state with at least two seats must prepare and approve a district map based on updated populations. In addition to the U.S. Constitution and federal law, congressional redistricting must comply with state redistricting laws and processes \cite{USredist}.

Attempts at enacting (and restricting) gerrymanders in the United States date back more than 200 years. There is not a specific legal standard for what a gerrymander is or isn't, and it has generally proved difficult for plaintiffs to persuade courts that a particular plan is illegal \cite{EG}.

While the general public may employ the ``eye test'' to mock districts with tortuous bounds, districts like that may be drawn specifically to comply with directives to connect communities of interest \cite{Cutting}. On the other hand, district maps with visually appealing districts may still be gerrymandered in favor of one political party through the use of packing and cracking. Packing refers to fitting as many voters from one party into as few districts as possible. The preferred candidate of the packed party is likely to win in the packed districts while the party's total voting strength is weakened across the whole state. Cracking spreads voters from a political party into as many districts as possible in order to dilute voting strength.

In this paper, we add to the algorithm in \cite{Cutting} to create 215 Congressional district maps for the state of Iowa based on different random initial seeds. The algorithm forces output maps to obey necessary requirements with regard to population size and contiguity. The algorithm aims to optimize compactness and is run without without regard to partisanship. We then can compare the actual enacted 2021 map to our distribution of random maps using a variety of metrics: does the actual map fall within the range of the random maps, or is it an outlier? 

An analysis of the vote share, efficiency gap, and compactness scores of the enacted map against the simulated maps suggest evidence of partisan bias during the redistricting process. While results from the enacted map are not outliers, it performs worse in the vote share in comparison with simulated maps.

\section{Redistricting Guidelines and Procedure}\label{S:redist} 



\subsection{Iowa Standards for Redistricting}

In Iowa, the redistricting process has been led by the nonpartisan Legislative Service Bureau since 1980 \cite{IALSB}. They are in charge of preparing an initial proposal and a potential second proposal, each of which the legislature can only approve or reject. If a third proposal is needed, then the legislature can begin to edit maps directly. 

\subsubsection{Equal apportionment}
The population in Iowa from the 2020 census was 3,190,369. The state was allocated four seats in the House of Representatives, so four districts need to be formed, each with an ideal population of 797,592.25. The Iowa Code, Section 42.4(1) requires districts to have a population deviation not to exceed 1\% of the ideal population \cite{IAcode}. This gives a population range from 789,616 to 805,568 for each district.

\subsubsection{Political boundaries}
A 1968 amendment to the Iowa Constitution \cite{IAconst} requires that individual counties not be divided into multiple congressional districts. This is relatively unique among US states: many states have at least one populous county which has more people than any single Congressional district. 
Even neighboring Nebraska, with three Congressional districts and less than 600,000 people in its largest county, splits two different counties between districts \cite{NEdist}. 

\subsubsection{Contiguity and compactness} The redistricting amendment \cite{IAconst} further requires that each district be compact and contiguous. The Article requires excluding maps where a district is only connected by a pair of counties which meet only at adjoining corners. 

The Iowa Code specifies that ``compactness" includes districts that are square, rectangular, or hexagonal in shape and excludes irregular shapes, barring any geographic or political boundaries. The code specifies two metrics, length-width compactness and perimeter compactness, to compare the relative compactness of two districts. The length-width compactness is computed by calculating the difference between the length and width, with a difference of zero signaling optimal compactness. The greater the perimeter compactness, the smaller the district perimeter \cite{IAcode}.

These metrics for compactness are somewhat dated. We iterate the shortcomings in Iowa's length-width compactness measure noted in \cite{PP}:
\begin{quote} \textit{
    Deviations that do not alter these four critical points have absolutely no effect on the district's score. The Iowa measure is thus blind to any connivance that occurs within the four ``walls" to the north, south, east, and west, although this may be very important to partisans.}
\end{quote} Polsby and Popper further remark that a district of the same shape may yield different length-width compact scores when rotated. In this paper, we will use more sophisticated metrics for evaluating the compactness of districts, introduced in Section \ref{SSS:compactness}

\subsection{Rejection of Proposal 1 in 2021}\label{ss:2021_redist}

In the 2021 redistricting process, the first map was rejected by the legislature and the second map was enacted \cite{AkinOct28}. While the initial set of maps was supported by the Democratic minority in the state legislature, the Republican majority claimed that population in the districts was unbalanced and that the shapes were insufficiently compact \cite{AkinOct5}. The legislature is not supposed to consider partisan balance or the location of incumbents when selecting maps, but news outlets also noted at the time that the first map created a likely Democratic First district. On the other hand, the second map split Linn and Johnson counties, both Democratic-leaning, between the First and Second districts, giving Republicans an advantage in both \cite{AkinOct28,DMR}.

\subsection{The Redistricting Algorithm}\label{SS:algorithm}

In this paper, we began by following the algorithm proposed by Chen and Rodden in \cite{Cutting} to generate 215 potential maps of Iowa Congressional Districts. A key component of this algorithm is to identify and work with the smallest geographical units of each congressional district. In Iowa, this is the 99 counties. The algorithm proceeds as follows:
\begin{enumerate}
    \item Identify each county as a separate district. There are now $i=99$ districts.
    \item Randomly select one of the $i$ districts. Join it to its nearest neighbor by calculating the distance between centroids. There are now $i-1$ districts
    \item Repeat Step 2 until there are four districts. These will be compact and contiguous, but may not have similar populations
    \item Find the pair of bordering districts with the largest gap in population. Call the smaller district $D_m$ and the larger $D_n$. Find all counties in $D_n$ which are adjacent to $D_m$ and which would not cause $D_n$ to become discontiguous if removed. Calculate the relative distance between each of these counties and the centroids of $D_n$ and $D_m$. Move the county with the highest relative distance from $D_n$ to $D_m$.
    \item Repeat Step 4 until the populations of all four districts are within your preset tolerance level. A county is ineligible for reallocation if
    \begin{enumerate}
        \item it has been switched more than five times or 
        \item if its population is six times greater than the population difference between the adjacent districts. 
    \end{enumerate}
    If a county is ineligible to be switched, step 3 will select another county.
\end{enumerate}

In each step, contiguity is preserved. By comparing the distance between the centroid of a particular county and the larger districts, we generally ensure that we minimize loss of compactness at each iteration of step 2 and maintain or improve compactness at each iteration of step 4.

\subsubsection{Improvements to the Algorithm for County Preservation}
The requirement to preserve county lines generally made the process of implementing the algorithm easier than if we had to deal with a far larger number of smaller geographies (townships, precincts, etc). However, it did present challenges in Step 4. If a county along the border or two districts with population disparities had an optimal position between the centroids of the two districts, the algorithm may repeatedly select that particular county for reallocation.

In order to account for this scenario, we added conditions referenced in Step 5 to the algorithm. First, we added a count which kept track of how many times individual counties were switched during the entire reallocation process. The switch count was assigned to each county through the entire reallocation process for a single map. 

Second, we set a county population of six times the population difference between two counties as a cut-off for a county's reallocation eligibility. This feature of the algorithm kicked into place as the populations in all districts approached the ideal population size. Since 11 of Iowa's 99 counties have populations less than 50,000 per the 2020 census, this condition still allowed most counties to be switched and reduced run-time by preventing the largest counties from switching districts.

This article focuses on Iowa as a case study and thus requires preservation of county lines. The conditions in Step 5 can be adjusted to preserve municipal boundaries rather than county lines.


\section{Comparison of Enacted and Simulated Maps}\label{S:analysis} 

Once we created the 215 maps\footnote{Duplicate maps were removed prior to analysis.}, we aimed to answer the following questions:
    \begin{enumerate}
        \item\label{analysis_goal1} How do the enacted and 2021 Proposal 1 maps rank in comparison with each other and the simulated maps?
        \item\label{analysis_goal2} Are the 2024 election results an outlier with respect to election results from the simulated maps using the same election data?
        \item\label{analysis_goal3} Based on 2024 election data, how do election results from simulated and enacted maps compare with election results from the last 40 years?
    \end{enumerate}

\begin{figure}[H]
\centering
\includegraphics[width=.8\textwidth]{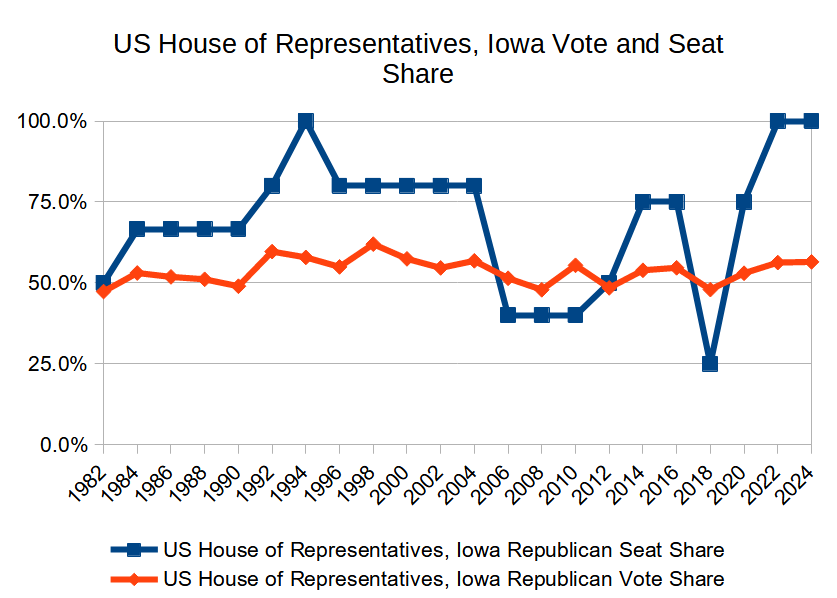}
\caption{Aggregated Vote Share and Seat Share over time}
\label{fig:VoteShareHist}
\end{figure}

In addressing point \ref{analysis_goal3}, we consider Iowa's context. Iowa has a history of voting for both Republicans and Democrats for statewide office. It voted for the Democratic nominee for president in six of the last 11 elections (1988, 1992, 1996, 2000, 2008, 2012), but voted for Donald Trump in the most recent three elections (2016, 2020, 2024) \cite{ElectionStats}. \autoref{table:VoteShareHist} and \autoref{fig:VoteShareHist} show the recent history of vote share and seat share for US House of Representatives elections from Iowa over the last five redistricting cycles. Vote share is calculated looking at only Republican and Democratic votes, ignoring third paries and write-ins.

\begin{table}[H]
\centering
\begin{tabular}{|c|c|c|c|c|c|}
\hline
  &&&R Seat&US House R&Pres. R\\
  & R Seats &D Seats &Share &Vote Share &Vote Share \\\hline\hline
1982 & 3 & 3 & 50\% & 47.3\% & \cellcolor[HTML]{C0C0C0} \\
1984 & 4 & 2 & 66.7\% & 53.1\% & 53.7\% \\
1986 & 4 & 2 & 66.7\% & 51.9\% & \cellcolor[HTML]{C0C0C0} \\
1988 & 4 & 2 & 66.7\% & 51.1\% & 44.9\% \\ 
1990 & 4 & 2 & 66.7\% & 49.0\% & \cellcolor[HTML]{C0C0C0} \\ \hline
1992 & 4 & 1 & 80\% & 59.7\% & 46.3\% \\ 
1994 & 5 & 0 & 100\% & 57.9\% & \cellcolor[HTML]{C0C0C0} \\ 
1996 & 4 & 1 & 80\% & 55.0\% & 44.3\% \\ 
1998 & 4 & 1 & 80\% & 62.0\% & \cellcolor[HTML]{C0C0C0} \\ 
2000 & 4 & 1 & 80\% & 57.4\% & 49.8\% \\ \hline
2002 & 4 & 1 & 80\% & 54.6\% & \cellcolor[HTML]{C0C0C0} \\ 
2004 & 4 & 1 & 80\% & 56.8\% & 50.3\% \\ 
2006 & 2 & 3 & 40\% & 51.5\% & \cellcolor[HTML]{C0C0C0} \\
2008 & 2 & 3 & 40\% & 47.9\% & 45.2\% \\ 
2010 & 2 & 3 & 40\% & 55.5\% & \cellcolor[HTML]{C0C0C0} \\ \hline
2012 & 2 & 2 & 50\% & 48.5\% & 47.0\% \\ 
2014 & 3 & 1 & 75\% & 53.9\% & \cellcolor[HTML]{C0C0C0} \\ 
2016 & 3 & 1 & 75\% & 54.7\% & 55.1\% \\ 
2018 & 1 & 3 & 25\% & 48.0\% & \cellcolor[HTML]{C0C0C0} \\ 
2020 & 3 & 1 & 75\% & 53.0\% & 54.2\% \\ \hline
2022 & 4 & 0 & 100\% & 56.3\% & \cellcolor[HTML]{C0C0C0} \\ 
2024 & 4 & 0 & 100\% & 56.5\% & 56.7\% \\ \hline
\end{tabular}
\caption{Seat Share and Vote Share for Iowa Elections for House of Representatives and US President, 1982-2024. Third party and write-in votes are ignored. Horizontal rules separate different decennial district maps \cite{ElectionStats}.}
\label{table:VoteShareHist}
\end{table}

\subsection{Analysis of the Enacted Congressional Map}

The reallocation process outlined in Subsection \ref{SS:algorithm} preserves county lines, maintains contiguity, and restricts individual population deviation to 1\% of the ideal district population. We focus on how well the enacted and alternate maps satisfy the compactness and equal apportionment criteria for redistricting.

\subsubsection{Compactness}\label{SSS:compactness}
Two measures to quantify the compactness of a region which are commonly utilized in redistricting work are the Reock and Polsby-Popper scores. The Reock score is the ratio of the area of the district to that of its minimum bounding circle \cite{Reock}. The Polsby-Popper score is the ratio of the area of the district to the area of a circle with the same perimeter as the district \cite{PP}. In both cases, the scores fall in the range $0<x\leq 1$, with a perfectly-compact circular district having a score of 1.

Generally, the overall compactness of a map is summarized either by looking at the minimum score (least-compact district) or the average score of all districts.


We computed the Reock and Polsby-Popper scores for each simulated map in addition to the two 2021 proposed maps. The enacted map is Proposal 2 and the alternate map is Proposal 1 from the 2021 redistricting process \cite{IA2021redist}. 


\begin{figure}[H]
\centering
\includegraphics[width=.75\textwidth]{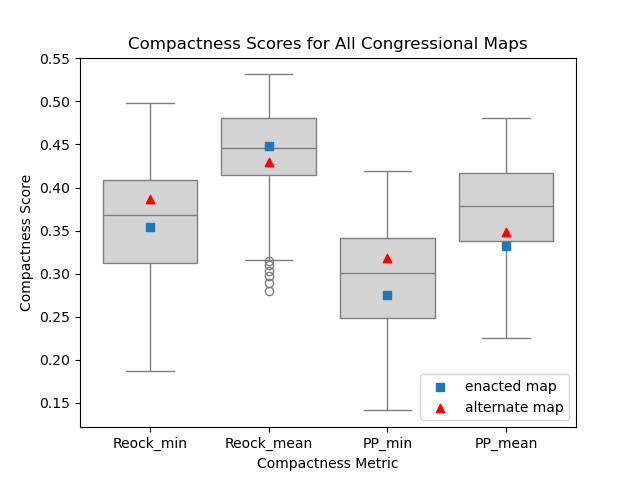}
\caption{Compactness score analysis for simulated, enacted, and alternate maps.}
\label{fig:compactnessbox}
\end{figure}


The enacted map ranks worse than the 2021 Proposal 1 map in three of four compactness metrics. 
Additionally, depending on the metric used, it is worse than 50\%-75\% of our computer-generated maps.




\subsubsection{Population}
In Table \ref{table:distpop}, the district populations of the enacted map and Proposal 1 maps are given. 
Republicans in the state legislature in 2021 claimed that the district populations were unbalanced \cite{AkinOct5}. However, the maximum population deviation are both less than 0.01\% and have no appreciable difference. The absolute mean district population deviation is 0.0033\% and 0.0039\% for the enacted and alternate maps, respectively.

\begin{table}[H]
\centering
\small
\begin{tabular}{|c|c|c|c|c|}
\hline
Map       & District 1  & District 2 & District 3 & District 4 \\ \hline\hline
Enacted   & 797,584     &  797,589      & 797,551              & 797,645          \\ \hline 
Alternate & 797,655      &  797,556     &   797,584            & 797,574             \\  \hline
\end{tabular}
\caption{Population by district for 2021 enacted and alternate maps.}
\label{table:distpop}
\end{table}

\begin{figure}[H]
\centering
\includegraphics[width=.75\textwidth]{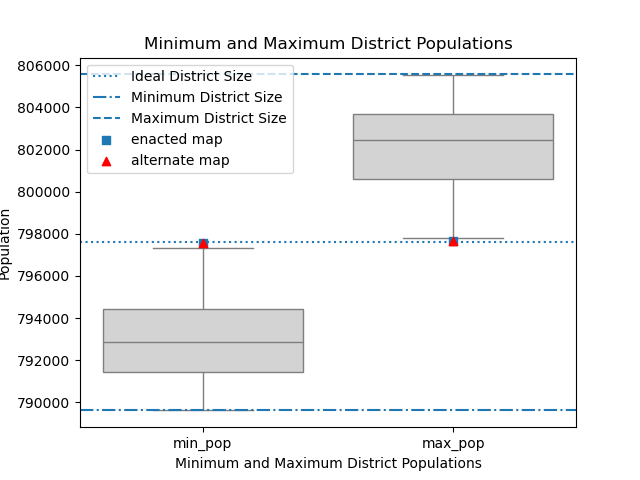}
\caption{Distribution of minimum and maximum district populations in simulated, enacted, and alternate maps.}
\label{fig:min_max_dist_pop}
\end{figure}

In \autoref{fig:min_max_dist_pop}, we plot the distribution of the largest and smallest district sizes for our 215 maps. Our algorithm was designed to terminate once all districts were within the 1\% threshold rather than attempt to optimize population further. 

\subsubsection{Remarks on the Rejection of Proposal 1}
The 2021 Proposal 1 map scores better than the enacted map in three out of four compactness score metrics. Additionally, in Figure \ref{table:pop_deviance}, we see that the enacted and Proposal 1 maps have negligible differences in their populations. The enacted map ranks between the 20th and 51st percentiles for the four compactness metrics given. The alternate map's compactness scores all fall within the interquartile range of the simulated maps, ranging from the 31st to 64th percentiles.
\begin{table}[H]
\centering
\begin{tabular}{|c|c|c|c|}
\hline
Map & Max deviance (\#)& Max deviance (\%) & Mean deviance (\%)\\ \hline\hline
Enacted  & 52.75& 0.0066\%  & 0.0033\% \\ \hline
Alternate & 62.75& 0.0079\%  & 0.0039\% \\ \hline
\end{tabular}
\caption{Deviance of enacted and alternate maps from ideal population of 797,592.25.}
\label{table:pop_deviance}
\end{table}

We find that the reasons cited by Republican legislators for rejecting Proposal 1 are not supported by the evidence provided by the random simulations. Using 2022 and 2024 election results, Proposal 1 map produces one Democratic and three Republican seats. The evidence suggests that Proposal 1 was rejected due to partisan bias.

\subsection{Vote Share}\label{ss:vote_share}

Having established that our 215 randomly-generated maps are similarly compact to the official map (and often more compact), we then proceed to evaluate the partisanship of the enacted and alternate maps against the simulated maps. For each of the four districts in each of the 215 maps, we totaled the Democratic and Republican votes for the 2024 presidential race\footnote{Section \ref{S:appendix} has analysis results for the 2022 congressional race.}

Note that we excluded third party votes as the mean-median difference and efficiency gap calculations use a two-party system. The actual turnout and total votes used in calculations are given in the following table.

\begin{table}[H]
\centering
\small
\begin{tabular}{|c|c|c|}
\hline
Election Year & Actual total & Two-party total \\ \hline\hline
2022          & 1,230,416    & 1,211,126       \\ \hline
2024          & 1,674,011    & 1,634,297       \\ \hline
\end{tabular}
\caption{Actual and two-party voter totals in presidential votes in 2024.}
\label{table:voter_total}
\end{table}

\begin{figure}[H]\label{rank_order_boxplot}
\centering\includegraphics[width=.75\textwidth]{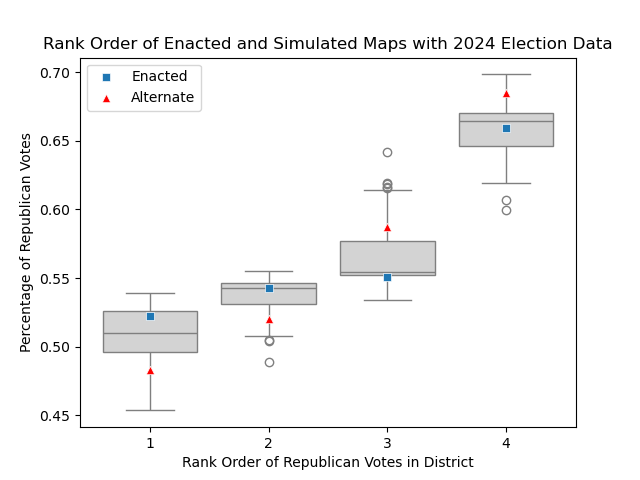}
\caption{Rank order boxplot of Republican vote share with 2024 election data across simulated, enacted, and alternate maps.}
\label{fig:ranking_box_2024}
\end{figure}

We then ranked the the four districts by percent of Republican voters, creating a monotonic vector. We plotted points representing the enacted and alternate maps on top of a box plot (in \autoref{fig:ranking_box_2024}) of the results for the 215 algorithmic maps. 


Visualizations like these are helpful in identifying maps which favor one political party over another. Since voter identification usually changes on some degree of a continuum, one might expect that the percentage of Republican voters might form a relatively smooth curve across different districts, especially for states with many districts.

In contrast, a graph like this could include a distinct jump for the most-Republican (or most-Democratic) district if packing has occurred. If the people drawing the maps are considering political ideology while constructing districts (which is not allowed under many state rules, including Iowa's), then there might be another discontinuity near the middle of the graph, where marginal districts with approximately 50-50 distribution of Democratic and Republican voters have instead been pushed to be more solidly partisan for one particular party.

Looking at the dot plots, there is a notable difference between the enacted and alternate maps: the alternate map sees almost a linear pattern of vote share across the four districts, while the enacted map is significantly more Republican in the left two districts (the least-Republican overall) and more Democratic in the right two districts (the most-republican overall).

\begin{figure}[H]
\centering
\includegraphics[width=.75\textwidth]{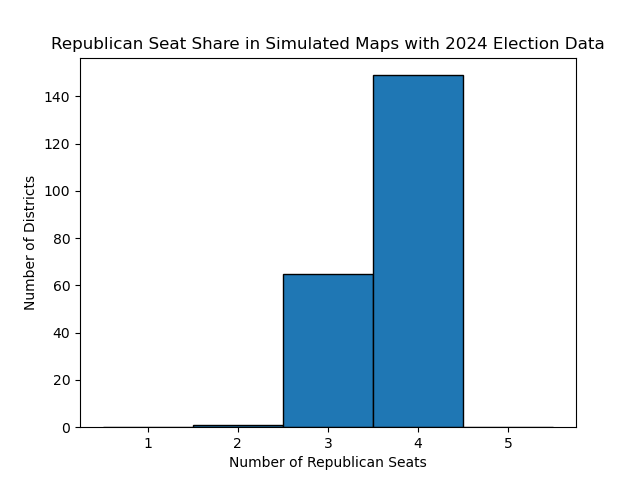}
\caption{The Republican seat share in simulated maps with 2024 election data.}
\label{rep_seatshare}
\end{figure}

\begin{figure}[H]
\centering
\includegraphics[width=.75\textwidth]{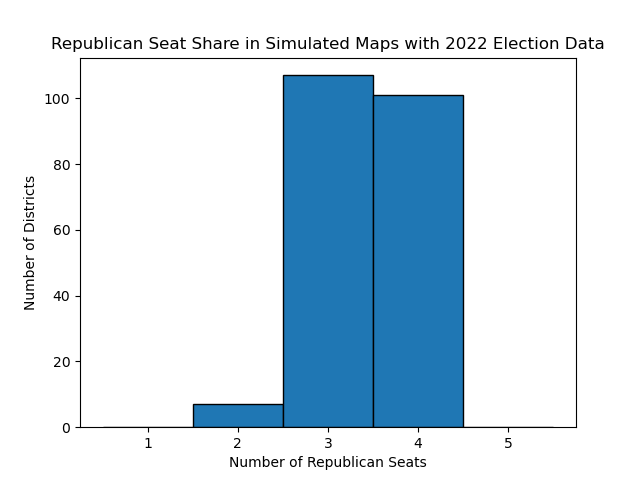}
\caption{The Republican seat share in simulated maps with 2022 election data. Note 56\% of maps have at least one Democratic seat.}
\label{fig:2022_rep_seatshare}
\end{figure}

Using 2024 presidential votes results, 30.7\% of simulated maps have at least one Democratic congressional seat. For 2022 congressional votes results, 53.0\% of simulated maps have at least one Democratic seat\footnote{See Figure \ref{fig:2022_rep_seatshare} in Section \ref{S:appendix} for 2022 results}.

\subsection{Efficiency Gap}
The efficiency gap is a numerical value that quantifies partisan bias in a two-party system. Votes cast for the losing party in a district are defined as lost votes. Any excess vote over the 50\% threshold for a party to win a district is defined as a surplus vote. A wasted vote is defined as the sum of lost and surplus votes over all districts. The efficiency gap is then computed by taking the difference between wasted votes for each party and dividing by the total number of votes cast statewide. An efficiency gap outside the range of -7\% to 7\% suggests that one party has an advantage with the enacted map \cite{EG}. 

We are interested in the effects of the enacted map and how the results compare to the distribution of efficiency gaps for the simulated maps and the efficiency gaps of elections since 2000.

We calculated the efficiency gap for each simulated map and the enacted map as outlined in \cite{EG} using 2024 presidential election results. The following scatterplot gives the efficiency gap against the number of Republican congressional seats. The upper right corner is concentrated as all Democratic votes are wasted votes when all congressional seats are lost. 

\begin{figure}[H]\label{efficiencygap}
\centering
\includegraphics[width=.75\textwidth]{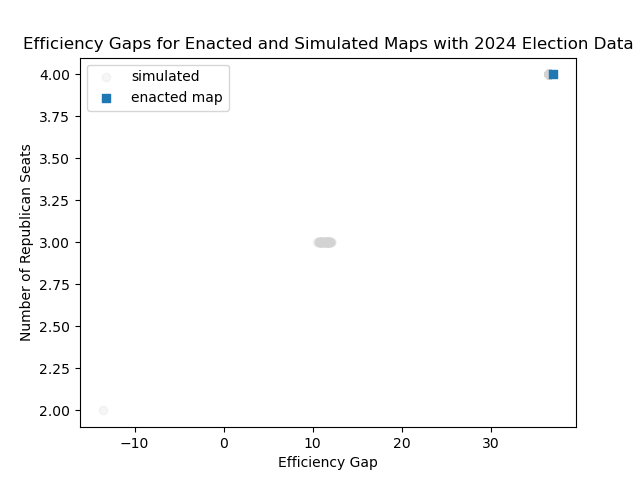}
\caption{Efficiency gaps of simulated maps with 2024 election data.}
\end{figure}

Roughly 69\% of simulated maps have four Republican congressional seats. One the one hand, the efficiency gap of the enacted map's 2024 election results falls within the interquartile range of distribution of simulated maps' efficiency scores. One the other hand, the 2024 election yields one of the largest efficiency gaps since 2000.

The following table shows the efficiency gap for U.S. House elections in Iowa since 2000 \cite{ElectionStats}. \begin{table}[H]
\centering
\begin{tabular}{|c|d{3}|}
\hline
     & \multicolumn{1}{|c|}{Efficiency Gap} \\ \hline\hline
2000 & $15.9\%$         \\ \hline
2002 & $19.4\%$        \\ 
2004 & $15.3\%$        \\ 
2006 & $-15.2\%$        \\ 
2008 & $-7.0\%$       \\ 
2010 & $-22.1\%$        \\ \hline
2012 & $2.4\%$          \\ 
2014 & $17.5\%$         \\ 
2016 & $15.8\%$         \\ 
2018 & $-22.0\%$        \\ 
2020 & $17.4\%$       \\ \hline
2022 & $37.4\%$         \\ 
2024 & $37.0\%$         \\ \hline
\end{tabular}
\caption{Efficiency Gap calculations for the last 13 elections of the US House of Representatives in Iowa. Horizontal rules separate different decennial district maps }
\label{table:effgap_hist}
\end{table}
None of these elections featured a candidate running unopposed, which can complicate calculations. In \autoref{table:effgap_hist}, we see that the efficiency gap has swung from favoring Republicans (positive gap) to favoring Democrats (negative gap) throughout this election cycle. Notably, the two largest efficiency gaps, by far, came in the 2022 and 2024 elections, with the current map.

\section{Conclusion}


Iowa's redistricting process, where initial maps are drawn by nonpartisan civil servants, is better than that of many other states. However, the maps are ultimately approved or rejected by a partisan legislature. While the legislature is not supposed to consider the partisan effects of the maps during the process, that does not mean that they are not aware of them. 

While voting trends in the last several election cycles trended more Republican \cite{ElectionStats}, the simulations show that it is not necessarily the case that we should expect four Republican congressional seats. As noted in Section \ref{ss:vote_share}, 30.7\% of simulated maps had one at least Democratic congressional seat. We present two maps with more optimal compactness scores than the enacted map. Simulated maps had a median score of 0.38 for minimum Reock score. The enacted map has a minimum Reock score of 0.35.

\begin{figure}[H]
\centering
\includegraphics[width=.75\textwidth]{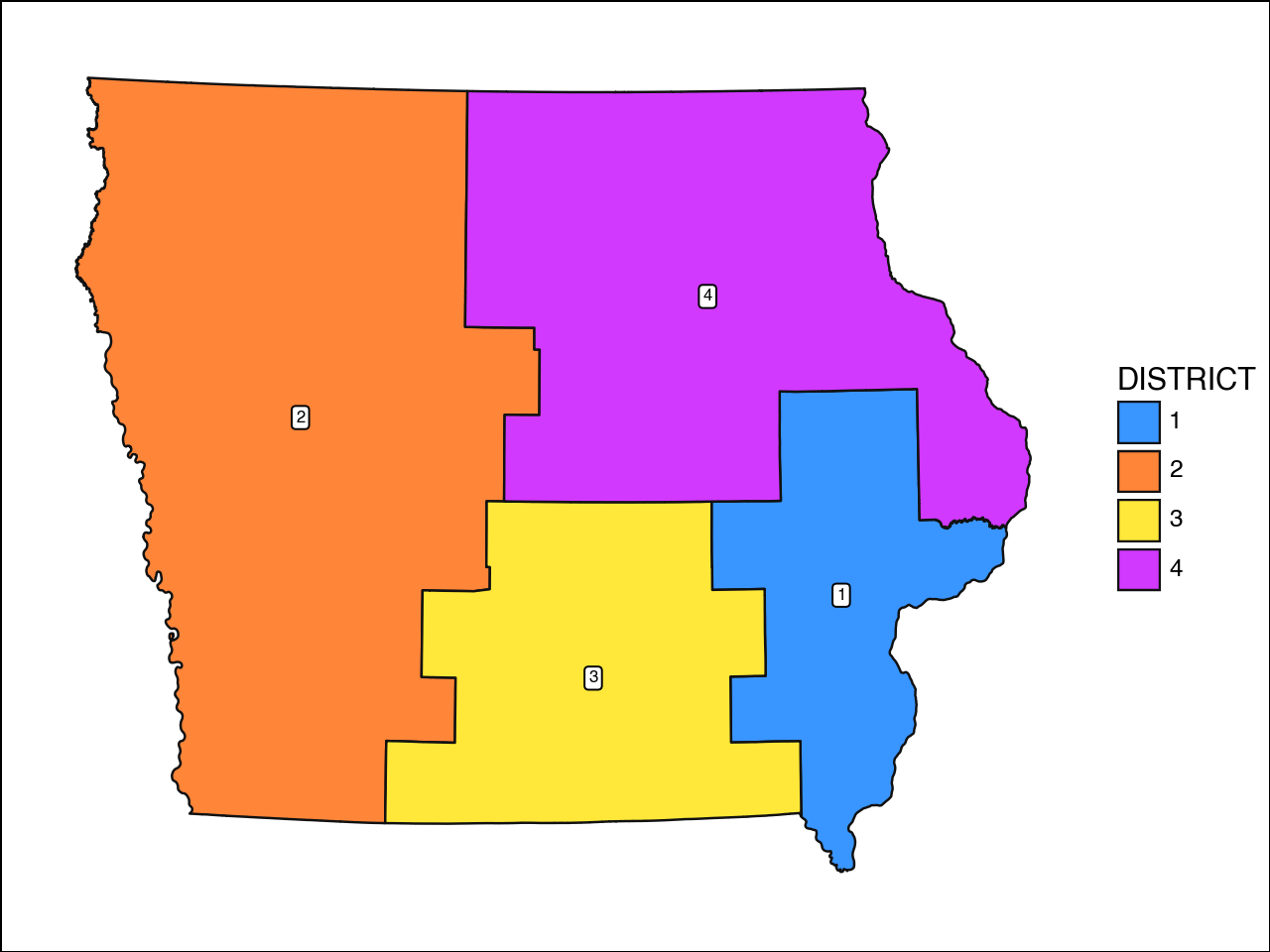}
\caption{A simulated map with a Reock minimum of 0.408 and one Democratic seat using 2024 election data.}
\label{fig:208_alt_map}
\end{figure}

The map in Figure \ref{fig:208_alt_map} has a minimum Reock score of 0.408 and absolute mean district population deviation from ideal population size of 0.66\%. With 2024 election results, the map in Figure \ref{fig:208_alt_map} yields one Democratic and three Republican congressional seats with an efficiency gap of 11.31\%.

The map in Figure \ref{fig:208_alt_map} has a minimum Reock score of 0.407 and absolute mean district population deviation from ideal population size of 0.22\%. With 2024 election results, the map in Figure \ref{fig:208_alt_map} yields four Republican congressional seats with an efficiency gap of 36.55\%.

\begin{figure}[H]
\centering
\includegraphics[width=.75\textwidth]{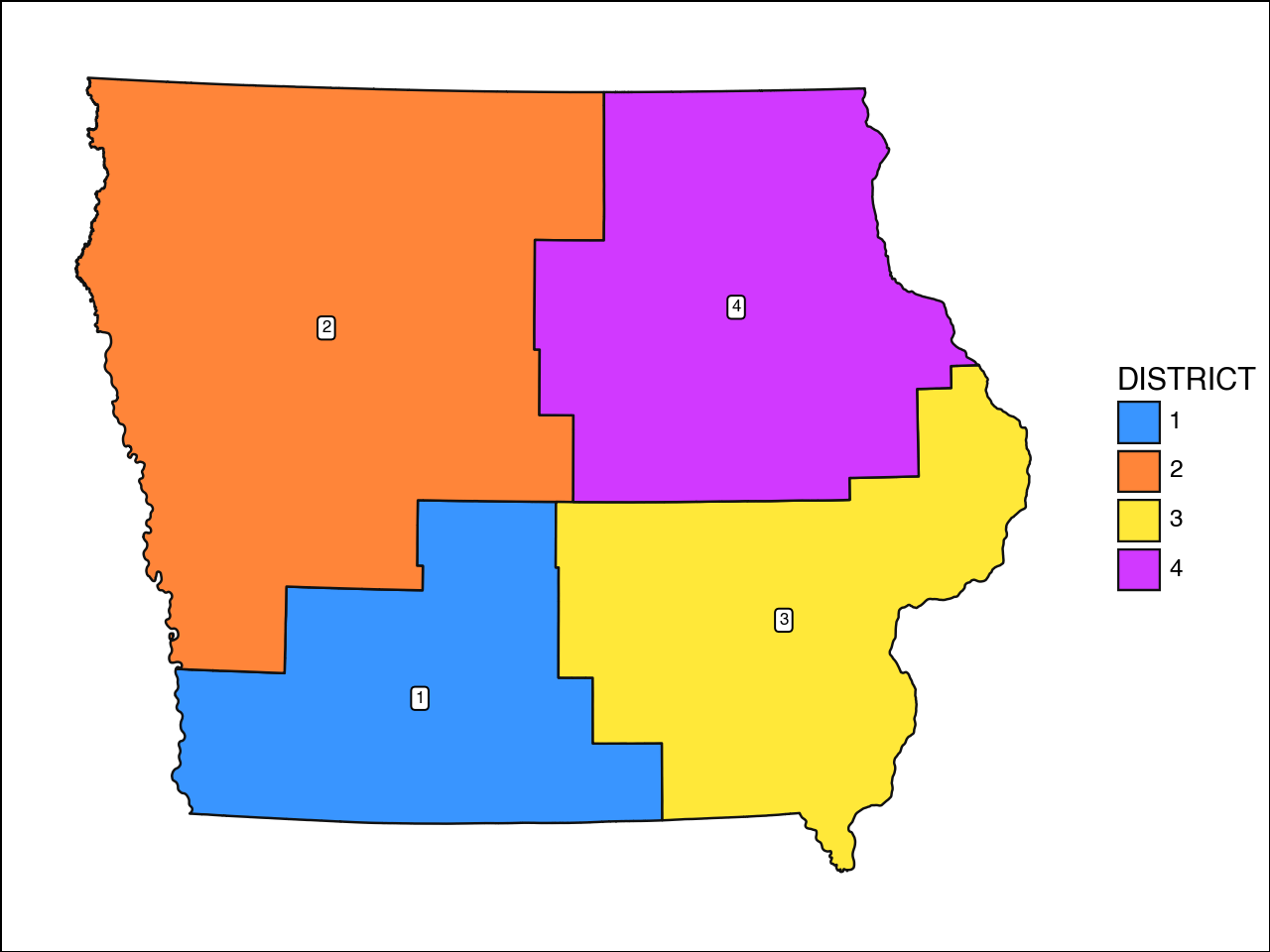}
\caption{A simulated map with a Reock minimum of 0.41 and no Democratic seats using 2024 election data.}
\label{fig:52_alt_map}
\end{figure}

Both maps rank better in terms of compactness than the enacted map. While the absolute mean district population deviation is much lower in the enacted map, the closest congressional races in 2022 and 2024 had a difference of thousands of votes between the winner and loser. No race was won by a matter of dozens of votes.

In examining the 2021 redistricting process, we find that both maps performed excellently on population distribution, perfectly on contiguity, and reasonably well on compactness. On the other hand, Proposal 1 was rejected for reasons poorly supported by the data.


\section{Appendix}\label{S:appendix}

\begin{figure}[H]\label{rank_order_boxplot}
\centering
    \includegraphics[width=.75\textwidth]{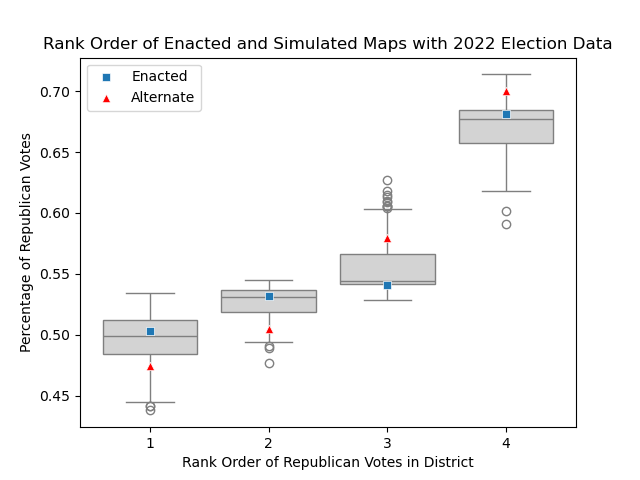}
    \caption{Rank order boxplot of Republican vote share with 2022 election data across simulated, enacted, and alternate maps.}
\label{fig:2022_ranking_box}
\end{figure}

\begin{thebibliography}{99}
\bibitem{DMR} ``{Editorial: Iowa redistricting can show again that partisanship need not be the only consideration in government.}'' \textit{Des Moines Register}, September 19, 2021.
\end{thebibliography}
\end{document}